\documentclass[aps,rmp,twocolumn,showpacs,a4paper]{revtex4}
\usepackage{graphicx}
 \usepackage{amsmath,amsfonts,amssymb}
\newcommand{\de}{\delta}

\newcommand{\f}{\phi}
\newcommand{\dd}{{\rm d}}

\renewcommand{\r}{{\bf r}}
\renewcommand{\i}{{i_1,i_2,i_3}}

\newcommand{\be}{\begin{equation}}
\newcommand{\ee}{\end{equation}}
\newcommand{\ba}{\begin{eqnarray}}
\newcommand{\ea}{\end{eqnarray}}
\newcommand{\nn}{\nonumber}
\newcommand{\qq}{\qquad}
\newcommand{\lb}{\label}
\begin{document}
\preprint{Submitted to Phys. Rev. Lett.}

%Title of paper
%\title{Poisson Solver using interpolating scaling functions}
\title{Efficient solution of Poisson's equation with free boundary conditions}

\author{Luigi Genovese, Thierry Deutsch}
\email{Thierry.Deutsch@cea.fr}
\affiliation{D\'epartement de recherche fondamentale sur la mati\`ere condens\'ee,\\
         SP2M/L\_Sim, CEA-Grenoble, 38054 Grenoble cedex~9, France}

\author{Alexey Neelov, Stefan Goedecker}
%\email[]{}
%\homepage[]{Your web page}
%\thanks{}
%\altaffiliation{}
\affiliation{Institute of Physics, University of Basel, Klingelbergstrasse 82, CH-4056 Basel, Switzerland}

\author{Gregory Beylkin}
\affiliation{Department of Applied Mathematics, University of Colorado at Boulder, Boulder, Colorado 80309-0526}

\date{\today}

 \begin{abstract}
Interpolating scaling functions give a faithful representation of a localized charge distribution by its 
values on a grid. For such charge distributions, using a Fast Fourier method, we obtain  highly accurate 
electrostatic potentials for free boundary conditions at the cost of O(N logN) operations, where N is the 
number of grid points. Thus, with our approach, free boundary conditions are treated as efficiently as 
the periodic conditions via plane wave methods.

 \end{abstract}

% insert suggested PACS numbers in braces on next line
%\pacs{ 111111111 }
% insert suggested keywords - APS authors don't need to do this
%\keywords{}

%\maketitle must follow title, authors, abstract, \pacs, and \keywords
 \maketitle

\section{Introduction}\label{Introduction}
Solving Poisson's equation
\begin{equation}\label{eqn:poisson}
\nabla^2 V=\rho\;,
\end{equation}
to find the electrostatic potential $V$ arising from a charge
distribution $\rho$ is a basic problem that can be found in nearly any
field of physics and chemistry. It is therefore essential to have efficient
solution methods for it.

A large variety of methods has been developed for systems of point particles
interacting by electrostatic forces. Formally this problem can be considered 
as the solution of Poisson's equation where the charge distribution is a sum 
of delta functions. The classical method for periodic boundary conditions is 
the Ewald method~\cite{ewald}. For large systems and free boundary conditions 
the Fast Multipole Method~\cite{FMM}  is a powerful method due to its linear 
scaling with respect to the number of particles. 

The FMM method has been generalized for continous systems where the charge density 
can be represented as a sum of Gaussians multiplied by a polynomial~\cite{scuseria}.
This generalization exhibits linear scaling with respect to the volume but not with 
respect to the number of Gaussians at constant volume. It works well in the context of 
quantum chemistry calculations with medium size Gaussian basis sets 
where the charge density is naturally obtained in the required form but 
it is not a general purpose method.

For periodic boundary conditions and smoothly varying charge densities 
plane wave  methods are simple and powerful because 
the Laplacian is a diagonal matrix in a plane wave representation. Given a Fourier 
representation of the charge density one can therefore obtain the potential simply 
by dividing each Fourier coefficient by $|{\bf k}|^2$, where ${\bf k}$ is the wavevector 
of the Fourier coefficient. If the charge density is originally given in real 
space a first Fast Fourier Transformation (FFT) is required to obtain its Fourier coefficients, 
and a second FFT is required to obtain the potential in real space from 
its Fourier coefficients. The overall scaling is therefore of order $N \log{N}$ where $N$ is the 
number of grid points. 

Because of the simplicity of these plane wave  methods various attempts have been made 
to generalize them to free boundary conditions. The most rudimentary method is just 
to take a very large periodic volume and to hope that the amplitude of the potential 
is nearly zero on the surface of the volume. Due to the long range of electrostatic 
forces this condition is however not fulfilled for volume sizes that are affordable 
with plane waves. Such as scheme is only possible if adaptive periodic wavelets 
are used~\cite{oleg}. 
In addition periodic boundary conditions do not permit to treat systems with monopoles 
and dipoles because for such systems no well defined solution exists under periodic 
boundary conditions. 
The first method to attack the problem in a systematic way was by Hockney~\cite{Hockney.1970}.
He proposed a Fourier approximation to the kernel 
\begin{equation}
\label{1-over-r}
K(r) = \frac{1}{r}
\end{equation}
of the Poisson equation. The method was intended for applications in plasma physics where 
no high accuracy is required.
In other application such as electronic structure calculations high accuracy is required 
and the method is not optimal. 
For a spherical geometry the Fourier coefficients of the $ \frac{1}{r} $ kernel can be 
calculated analytically. This is the basis of the simple and powerful method by 
F\"{u}sti-Molnar and Pulay~\cite{Fusti-Molnar-Pulay.2002}. Its obvious restriction is that it 
is efficient only for spherical geometries. 
Another rather complicated method was proposed by Martyna and Tuckerman. This method gives 
high accuracy only in the center of the computational volume. It requires therefore 
artificially large simulation boxes which is numerically very expensive. 
All the above discussed methods use FFT's at some point and have therfore  
a $ O(N \log N) $ scaling. 

In this paper we will describe a new Poisson solver for free boundary
conditions on an uniform mesh.  Contrary to Poisson solvers based on plane wave
functions, our method is using interpolating scaling functions to represent the charge density. 
It is therefore from the beginning free of long range interactions between
supercells, that falsify results if plane waves are used to describe non-periodic systems. 
Due to the convolutions we have to evaluate our method  has a $N \log(N)$ scaling 
instead of the ideal linear scaling,
Due to its small prefactor the method is however most efficient 
when dealing with localized 
densities such as can be found for example in the context of {\it ab initio} 
pseudo-potential electronic structure calculations using finite differences~\cite{chelikowsky} 
finite elements~\cite{pask} or plane waves for non-periodic systems.

\section{Interpolating scaling functions}
\begin{figure}
\includegraphics[width=.45\textwidth]{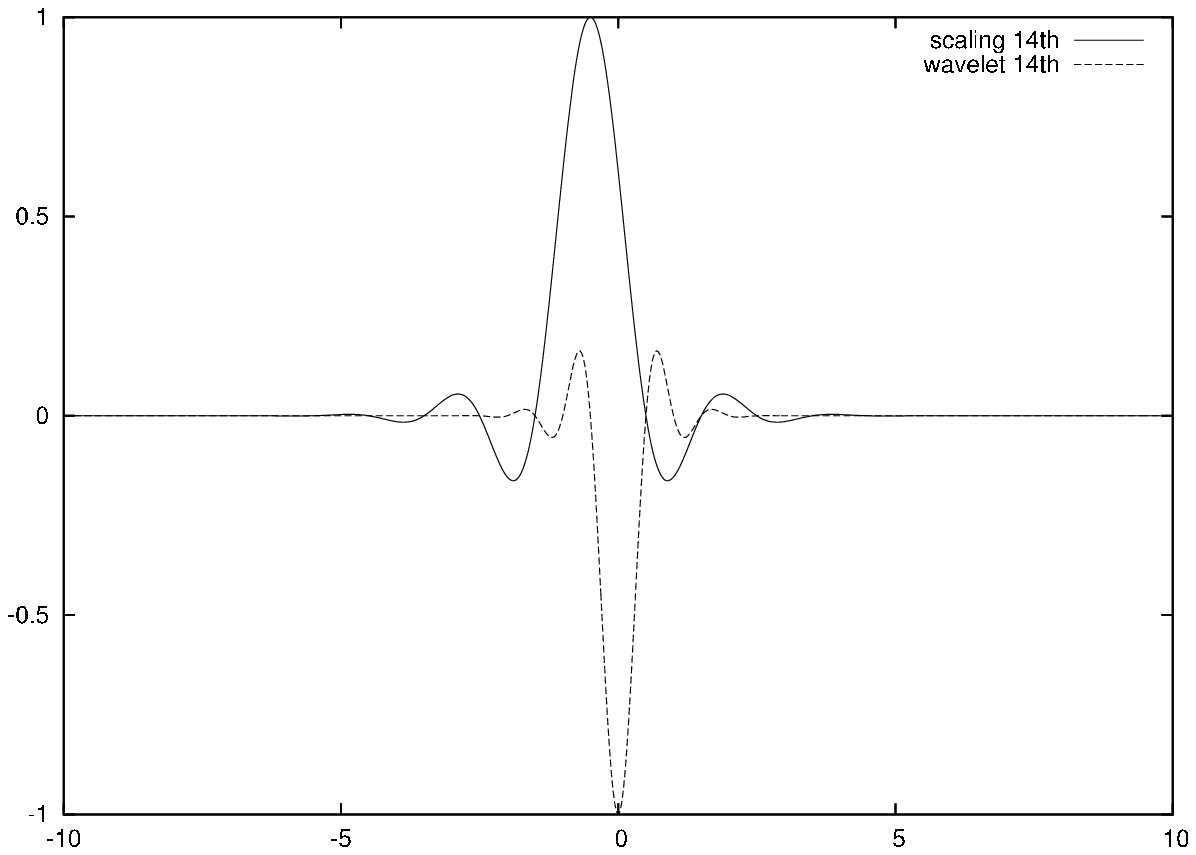}
\includegraphics[width=.45\textwidth]{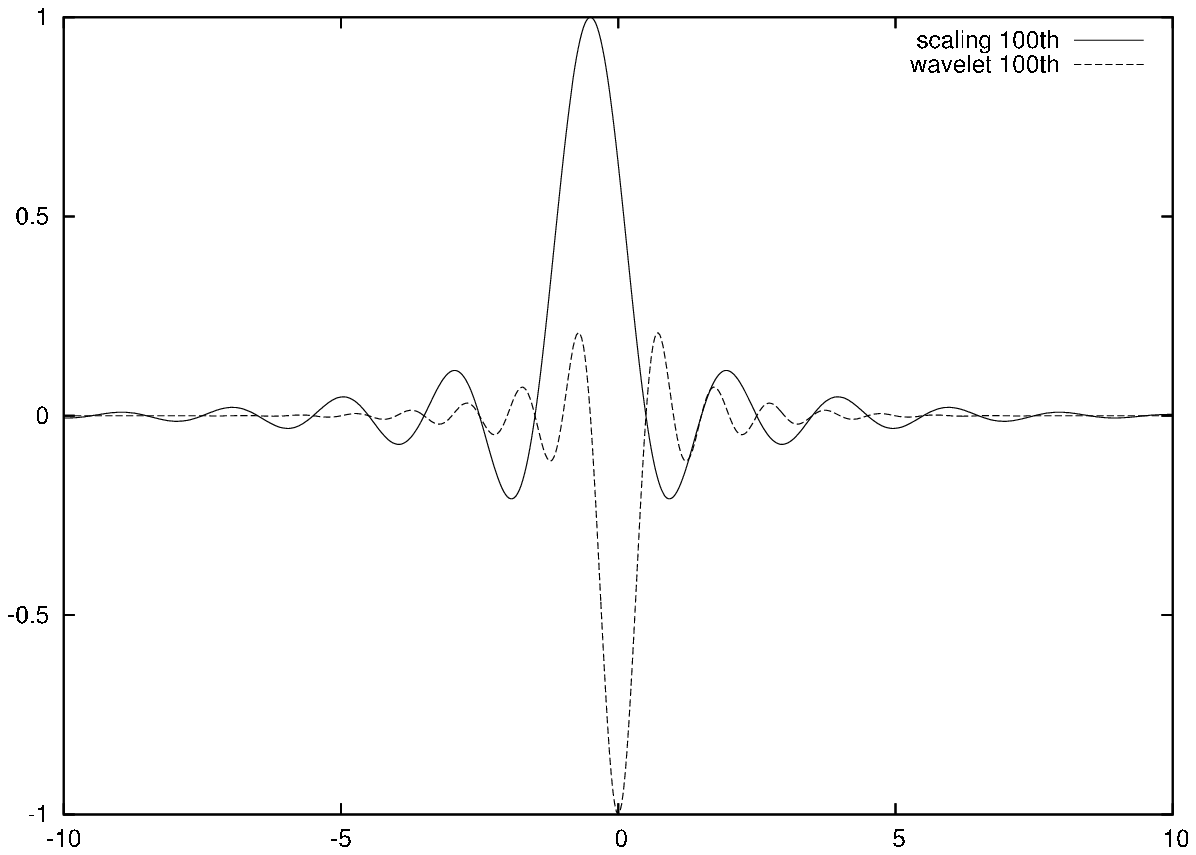}
\caption{Plots of interpolating scaling functions and wavelets of 14-th and 100-th order.} 
\label{iposcf}
\end{figure}

Scaling functions arise in wavelet theory ~\cite{daub}.
A scaling function basis set can be obtained from all the translations 
by a certain grid spacing $h$ of the mother wavelet centered at the origin. 
What distinguished scaling functions from other localized basis functions like 
finite elements is the refinement relation. The refinement relation 
establishes a relation between a scaling function $\phi(x-i)$ and 
the same scaling functions compressed by a factor of two, or,  
equivalently, between the scaling functions on a grid with grid spacing $h$ and 
another one with spacing $h/2$. 
For the scaling function centered a the origin, it reads 
\begin{equation}
\label{refinement}
\phi(x) = \sum_{j=-m}^{m} h_j \: \phi(2 x -j)
\end{equation}
where the $h_j$'s are the elements of a filter that characterizes the wavelet family. 
An interpolating wavelet has in addition the property that it is equal to one at the 
origin and zero at all other integer points. Because of this property it is very simple 
to find the scaling function expansion coefficients of any function. The coefficients 
are just the values of the function to be expanded on the grid. 
$m$-th order interpolating scaling functions are generated by $m-1$-th order 
recursive interpolation~\cite{lazy}. 
Fig.~\ref{iposcf} shows an 14-th order and 100-th order interpolating scaling function. 
Three-dimensional scaling functions can be obtained as the product of 
their one-dimensional counterparts
\begin{align}
\label{Phi-localized}
\Phi_{i_1,i_2,i_3}({\bf r}) &= \phi_{i_1}(x) \: \phi_{i_2}(y) \: \phi_{i_3}(z)\notag \\
 &= \phi(x-i_1) \: \phi(y-i_2) \: \phi(z-i_3)
\end{align}
where ${\bf r} = (x,y,z)$. The points
$i_1 , i_2 , i_3$ are the nodes of an uniform 3-dimensional mesh, with $i_p=1,\cdots,n_p$, $p=1,2,3$.

Continuous charge distributions are represented in numerical work typically by 
their values $\rho_{i,j,k}$ on a grid. It follows from the above described properties 
of interpolating scaling functions that the corresponding continous charge 
density is given by 
\begin{equation}
\label{rhomap}
\rho({ \bf r}) = \sum_{i_1,i_2,i_3} \rho_{i_1,i_2,i_3} \phi(x-i_1) \: \phi(y-i_2) \: \phi(z-i_3)
\end{equation}
The mapping of Eq.~\ref{rhomap} between the discretized and continous charge 
distribution ensures that the first $m$ discrete  and continous moments are
identical for a $m$-th order interpolating wavelet family, i.e. 
\begin{equation}
\sum_{i,j,k} i^{\ell_1} j^{\ell_2} k^{\ell_3} \: \rho_{i,j,k} =
 \int \dd {\bf r} \: x^{\ell_1} y^{\ell_2} z^{\ell_3} \: \rho({ \bf r})\;,\lb{theo}
 \end{equation}
 if $\ell_1,\ell_2,\ell_3 < m$.
The proof of this relation is given in the Appendix. 
Since the various multipoles of the charge distribution determine the major 
features of the potential the above equalities tell us that a scaling function 
representation gives the most faithful mapping between a continuous and discretized 
charge distribution for electrostatic problems. 

\section{Poisson's equation in a basis set of interpolating scaling functions}
As is well known the following integral equation gives the potential for free 
boundary conditions
\begin{equation}
 V({\bf r}) = \int \dd {\bf r}' \frac{1}{|{\bf r}-{\bf r}'|} \rho({\bf r}')\;.
\end{equation}
We are interested in the values of the potential on the same grid that was used 
for the charge density. Denoting the potential on the grid point 
${\bf r}_{j_1,j_2,j_3} = (x_{j_1}, y_{j_2}, z_{j_3}) $ by $ V_{j_1,j_2,j_3}=V({\bf r}_{j_1,j_2,j_3})$ 
we have
\begin{align}
V_{j_1,j_2,j_3} &=\\&= \sum_{i_1,i_2,i_3} \rho_{i_1,i_2,i_3}\int \dd {\bf r}' 
\frac{\phi_{i_1}(x') \: \phi_{i_2}(y') \: \phi_{i_3}(z')}{|{\bf r}_{j_1,j_2,j_3}-{\bf r}'|} \;.\notag
\end{align}
The above integral defines the discrete kernel
\begin{multline}
K(i_1,j_1;i_2,j_2;i_3,j_3) =\\= \int \dd {\bf r}' \phi_{i_1}(x') \: \phi_{i_2}(y') \: \phi_{i_3}(z')  
\frac{1}{|{\bf r}_{j_1,j_2,j_3}-{\bf r}'|} \;.
\end{multline}
Since the problem is invariant under combined translations of both the source point 
$(i_1,i_2,i_3)$ and the observation point $(j_1,j_2,j_3)$ the kernel depends only on the 
difference of the indices
\begin{equation}
K(i_1,j_1;i_2,j_2;i_3,j_3) = K(i_1-j_1,i_2-j_2,i_3-j_3)
\end{equation}
and the potential $V_{j_1,j_2,j_3}$ can be obtained from the charge density $\rho_{i_1,i_2,i_3}$ 
by the following 3-dimensional convolution:
\begin{equation}
V_{j_1,j_2,j_3} = \sum_{i_1,i_2,i_3} K(i_1-j_1,i_2-j_2,i_3-j_3) \rho_{i_1,i_2,i_3}\;.
\end{equation}
Once the Kernel is available in Fourier space, 
this convolution can be evaluated with two FFTs at a cost of $O (N \log N)$ operations
where $N = n_1 \: n_2 \: n_3 $ is the number of 3-dimensional grid points.
Since all the quantities in the above equation are real, real-to-complex FFT's 
can be used to reduce the number of operations compared to the case where 
one would use ordinary complex-complex FFT's. 
Obtaining the Kernel in Fourier space from the Kernel $K(j_1,j_2,j_3)$ in real space 
requires another FFT. 

It remains now to calculate the values of all the elements of the kernel $K(k_1,k_2,k_3)$. 
Solving a 3-dimensional integral for each element would be much too costly and we use 
therefore a separable approximation of $1/r$ in terms of 
Gaussians~\cite{Beylkin-Mohlenkamp.2002,harrison}, 
\begin{equation}
\label{beylkin}
\frac{1}{r} \simeq \sum_k \omega_k e^{-p_k r^2}\;.
\end{equation}
In this way all the complicated 3-dimensional integrals become products of simple 
1-dimensional integrals. 
Using $89$ Gaussian functions with the coefficients $\omega_k$ and
$p_k$ suitably chosen, we can approximate $\frac{1}{r}$ with an error
less than $10^{-8}$ in the interval $[10^{-9}\,, 1 ]$.
If we are interested in a wider range, e.g. a variable $R$ going from zero to $L$, we can use $r=\frac{R}{L}$:
\begin{eqnarray}
\frac{L}{R} & = & \sum_k \omega_k e^{-\frac{p_k}{L^2} R^2}\;,\\
\frac{1}{R} & = & \frac{1}{L}  \sum_k \omega_k e^{-P_k R^2}\;,\\
P_k & = & \frac{p_k}{L^2}\;.
\end{eqnarray}
With this approximation, we have that
\begin{equation}
K_{j_1,j_2,j_3} =  \sum_{k=1}^{89} \omega_k K_{j_1}(p_k) K_{j_2}(p_k) K_{j_3}(p_k)\;,
\end{equation}
where 
\begin{eqnarray}
K_j(p_k)	& = & \int \varphi_{j}(x) e^{-p_k x^2} \dd x\\
	& = & \int \varphi_{0}(x) e^{-p_k (x-j)^2} \dd x\;.
\end{eqnarray}
So we only need to evaluate $89$ $\times$ $\max\left(\{n_1,n_2,n_3\}\right)$ integrals of the type 
\begin{equation}
\label{kone}
K_j(p)  = \int \varphi_{0}(x) e^{-p (x-j)^2} \dd x\;,
\end{equation}
for some value of $p$ chosen between $3\cdot10^{-5}$ and
$3\cdot10^{16}$.

The accuracy in calculating the integrals can be further improved by
using the refinement relation for interpolating scaling functions~\ref{refinement}.
%\begin{equation}
%\varphi_0(x/2)  = \sqrt{2}\sum_j h_j \varphi_0(x-j) =\sqrt{2} \sum_j h_j \varphi_j(x)\;,
%\end{equation}
%where $h_j$ are the filters of the refinement relations of the functions $\{\varphi_i\}$

\begin{figure}
\includegraphics[width=.45\textwidth]{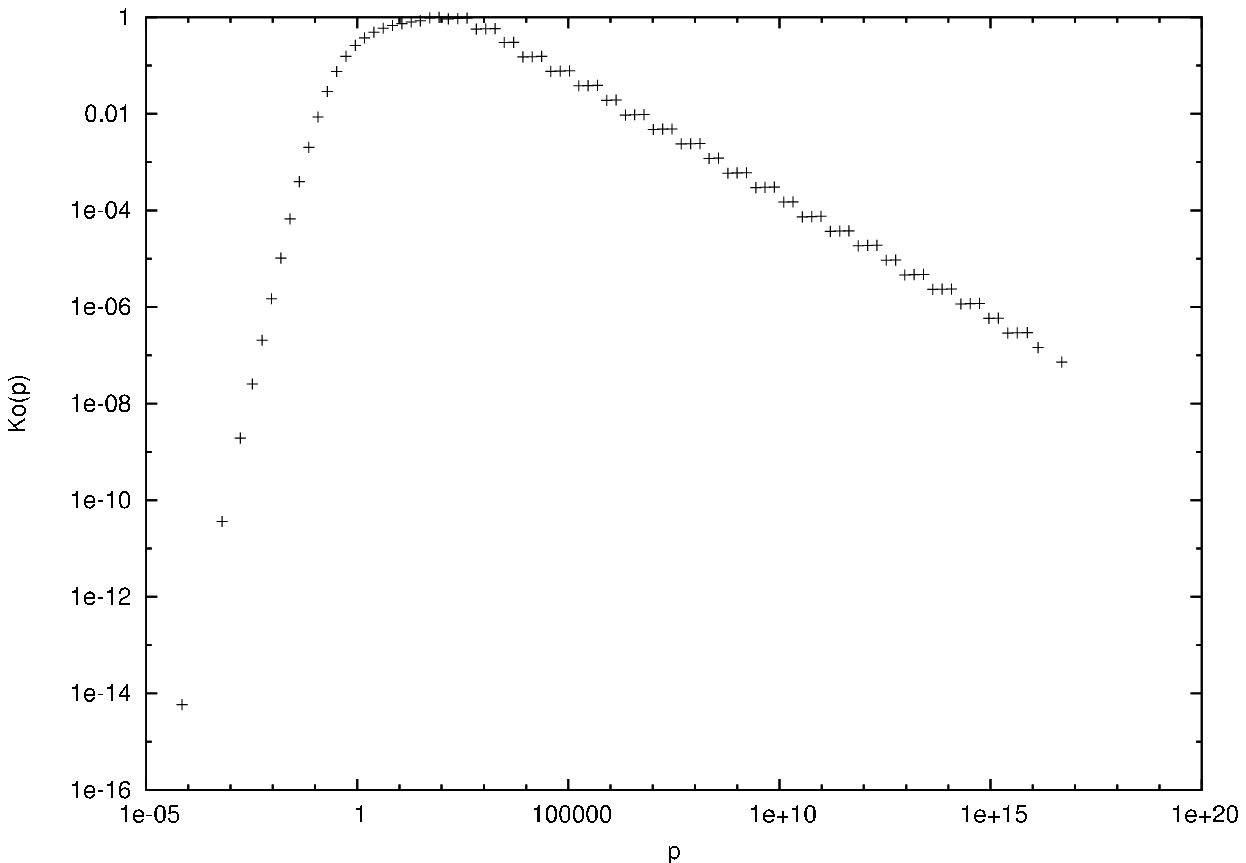} %gaussian-integral.pdf}
\includegraphics[width=.45\textwidth]{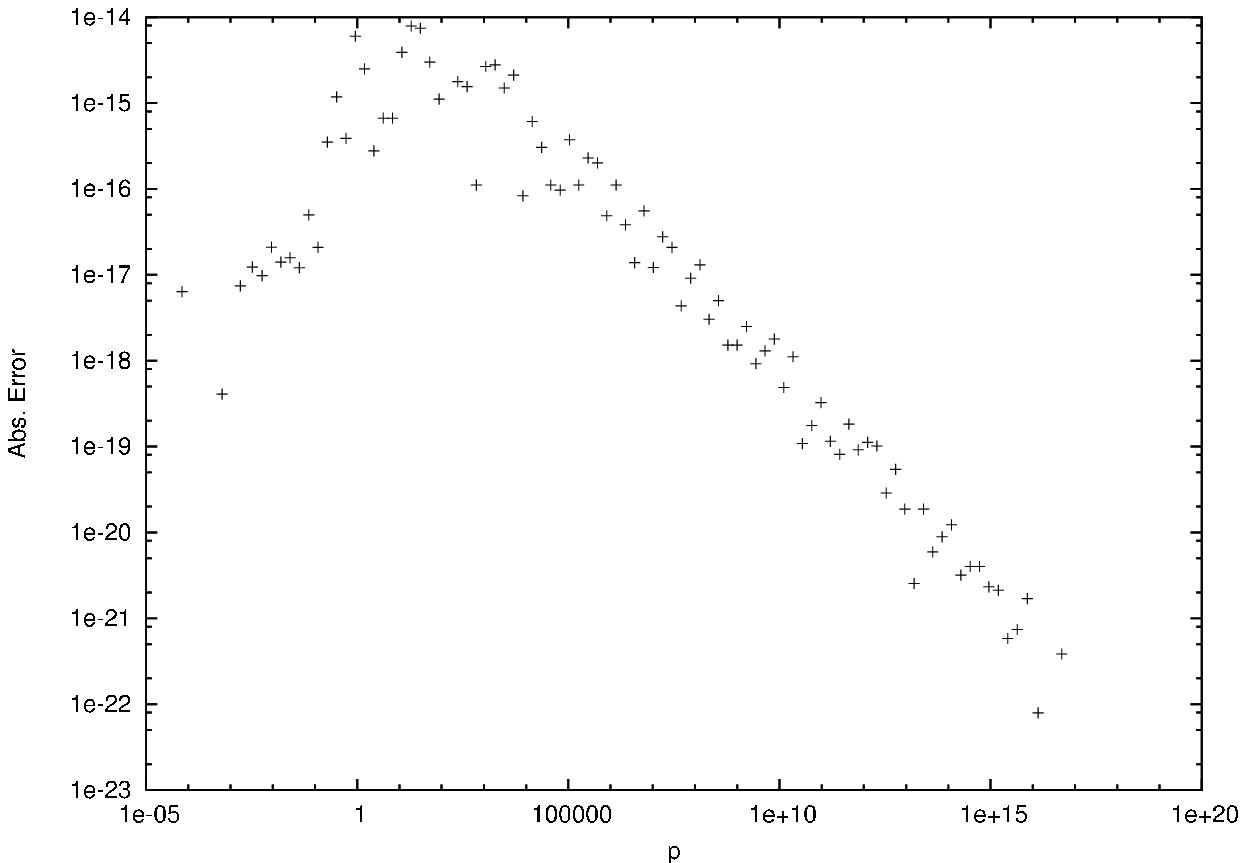}%diff-gaussian-integral.pdf}
\caption{Plots of the value of $K_0(p)$ (top) and error of the integration defining its value (bottom) of for 1024 integration points for all the values of $p$ used in the tensor decomposition of $1/r$ in gaussian functions.} 
\label{figure-gaussian-integral}
\end{figure}

From~\ref{kone}, we can evaluate $K_i(4p)$ as:
\begin{eqnarray}
K_i(4p) & = & \int \varphi(x) e^{-4p (x-i)^2} \dd x \\
        & = & \frac{1}{2} \int \varphi(x/2) e^{-p (x - 2i)^2} \dd x\\
        & = & \frac{1}{2} \sum_j  h_j
	      \int \varphi_j(x) e^{-p (x-2i)^2} \dd x\\
	& = & \frac{1}{2} \sum_j h_j K_{2i-j}(p)\;.
\label{recursion}
\end{eqnarray}
The best accuracy in evaluating numerically the integral is attained
for $p < 1$. 
For a fixed value of $p$ given by Eq.~\ref{beylkin}, 
the relation~\ref{recursion} is iterated $n=[\log_4(p)]$ 
times starting with $p_0=\frac{p}{4^n}$. So the
numerical calculation of the integrals $K_i(p)$ is performed as follows: for
each $p$, we compute the number $n$ of required recursions levels and
calculate the integral $K_i(p_0)$. The value of $n$ is chosen such
that $p_0 \simeq 1$ so we have a gaussian functions not too sharp. The
evaluation of the interpolating scaling functions is fast on a uniform
grid of points so we perform a simple summation over all the grid points. In
Figure~\ref{figure-gaussian-integral}, we show that $1024$
points are enough to obtain machine precision. 
Note that the values of $K_0(p)$ vary over many orders of magnitude as shown in 
In Figure~\ref{figure-gaussian-integral}. 

%L_Sim/BigDFT/PSolver-RealKernel/LAZY8

% \begin{figure}
% \label{figure-gaussian-integral}
% \caption{Value of the integral $K_0(p)$ for all the values of $p$ used in the tensor decomposition of $1/r$ in gaussian functions} 
% \end{figure}

\section{Numerical results and comparison with other methods}
\begin{figure}
\includegraphics[width=.33\textwidth,angle=-90]{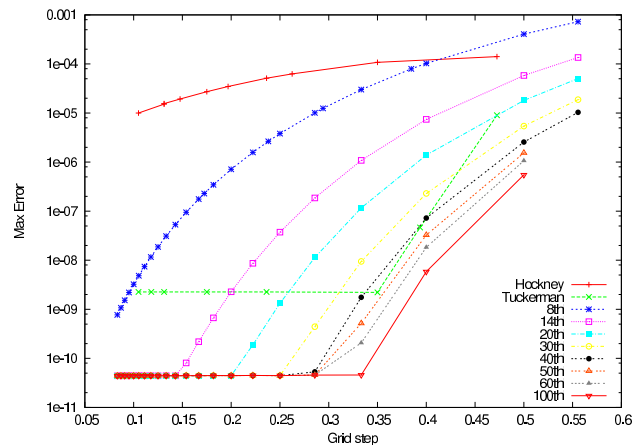}
\caption{Accuracy comparison between our method with interpolating scaling functions of different orders 
and the Hockney of Martyna-Tuckerman method as implemented in CPMD. The accuracy of our method is finally 
limited by the accuracy of the expansion of Eq.~\ref{beylkin} with 89 terms.} 
\label{comparison-methods}
\end{figure}

We have compared our method with the plane wave methods by Hockney~\cite{Hockney.1970}
and Martyna and Tuckerman~\cite{Martyna-Tuckerman.1999} as implemented in the CPMD 
electronic structure program~\cite{cpmd}. As expected Hockney's method does not 
allow to attain high accuracy. The method by Martyna and Tuckerman has a rapid exponential 
convergence rate which is characteristic for plane wave methods. Our new method has an algebraic 
convergence rate of $h^m$ with respect to the grid spacing $h$. By choosing very high order 
interpolating scaling functions we can get arbitrarily high convergence rates. Since convolutions 
are performed with FFT techniques the numerical effort does not increase as the order $m$ is increased. 
The accuracy shown in Figure~\ref{comparison-methods} for the Martyna and Tuckerman method is the 
accuracy in the central part of the cube that has 1/8 of the total volume of the computational cell. 
Outside this volume 
errors blow up. So the main disadvantage of this method is that a very large computational volume 
is needed in order to obtain accurate results in a sufficiently large target volume. 
For this reason the less acurate Hockney method is generally prefered in the CPMD program~\cite{hutter}. 

A strictly localized charge distribution, i.e. a charge distribution that is exactly zero outside 
a finite volume, can not be represented by a finite number of plane waves. This is an inherent 
contradiction in all the plane wave methods for the solution of Poisson's equation under free 
boundary conditions. For the test shown in 
Figure~\ref{comparison-methods} we used a Gaussian charge distribution whose potential 
can be calculated analytically.  The Gaussian was embedded in 
a computational cell that was so large that the tails of the Gaussian were cut off at an 
amplitude of less than 1.e-16. A Gaussian can well be represented by a relatively small number 
of plane waves and so the above described problem is not important. For other localized charge distributions 
that are less smooth a finite Fourier representation is worse and leads to a spilling of the 
charge density out of the original localization volume. This will lead to inaccuracies in the potential.

Table~\ref{timings} shows the required CPU time for a $128^3$ problem as a function of the number of processors on a Cray parallel computer. The parallel version is based on a 
parallel 3-dimensional FFT. 

%\clearpage
\begin{table}[h]
\begin{tabular}{|c|c|c|c|c|c|c|} \hline
  1 & 2 & 4 & 8 & 16 & 32 & 64  \\ \hline \hline
 .92 & .55 & .27 & .16 & .11 & .08 & .09 \\ \hline 
%%1.48 & .87 & .48 & .32 & .21 & .15 &  \\ \hline 
\end{tabular}
\caption[]{ The elapsed time in seconds required on a Cray XT3 (based on AMD Opteron processors) 
to solve Poisson's equation 
on a 128$^3$ grid as a function of the number of processors. Since Poisson's equation 
is typically solved many times, the time for setting up the Kernel is not included.
Including the set up time of the Kernel increases the total timing by about 50 percent, since  one 
additional FFT is needed.
\label{timings}}

\end{table}

A package for solving Poisson's equation according to the method described here can be downloaded from 
www.unibas.ch/comphys/comphys/SOFTWARE

In conclusion, we have presented a method that allows to obtain the potential of a localized 
charge distribution under free boundary conditions with a $O(N \log N)$ scaling in a mathematically 
clean way. Even though our method 
has the same scaling behaviour as existing plane wave methods it is not a plane wave method 
in the sense that neither the charge density nor the potential are ever represented by plane waves. 
Instead interpolating scaling functions are used for the representation of the charge density. 

We acknowledge interesting discussions with Reinhold Schneider and Robert Harrison.
This work was supported by the European Commission within the Sixth Framework Programme through 
NEST-BigDFT (contract no. BigDFT-511815) and by the Swiss National Science Foundation.
Research of G.B. was partially supported by DOE grant DE-FG02-03ER25583,
DOE/ORNL grant 4000038129 and DARPA/ARO grant W911NF-04-1-0281. 
The timings were performed at the CSCS (Swiss Supercomputing Center) in Manno.

% The Appendices part is started with the command \appendix;
% appendix sections are then done as normal sections
\appendix
\section{}

In the present Appendix we are going to prove Eq. (\ref{theo}):
 Let $\f(x)$ be an interpolating scaling function of Deslariers-Dubuc, of the order $m$, and  $\rho_\i$ be a three-dimensional array of constant coefficients. Let, further, 
\ba
\rho(\r)=\sum_\i \rho_\i \f(x-i_1)\f(x-i_2)\f(x-i_3).\lb{rho}
\ea
Then, 
\ba
\sum_\i i_1^{l_1}i_2^{l_2}i_3^{l_3}\rho_\i=\int d\r x^{l_1} y^{l_2} z^{l_3}\rho(\r) \lb{theor}\\
{\rm if}\qq 0\leq l_1,\,l_2,\,l_3<m\nn
\ea

This follows from the fact, proven in reference~\cite{beyl} that the first $m$ moments of the scaling function obey the formula:
\ba
M_l=\int \f(x) x^l \dd x=\de_l,\qq l=0,..,m-1 \lb{moms}
\ea
Shift the integration variable, we have
\ba
&&\int \f(x-j) x^l \dd x=\int \f(t) (t+j)^l \dd t=\nn\\
 &=&
\int \f(t)\sum_{p=0}^l C_l^p t^pj^{l-p}\dd t=j^l\nn
\ea
Then, inserting (\ref{rho}) into the right side of (\ref{theor}), we get:
\ba
&&\int \dd\r x^{l_1} y^{l_2} z^{l_3}\rho(\r)=
\int  x^{l_1} y^{l_2} z^{l_3}
\sum_\i \rho_\i\times\nn\\ 
&\times& \f(x-i_1)\f(x-i_2)\f(x-i_3)\dd\r=\sum_\i \rho_\i \times\nn\\
&\times& \int x^{l_1}\f(x-i_1)\dd x 
\int y^{l_2}\f(y-i_2)\dd y\times\nn\\
&\times& \int z^{l_3}\f(z-i_3)\dd z=\sum_\i \rho_\i i_1^{l_1} i_2^{l_2} i_3^{l_3}\nn
\ea

%\bibliography{../Article}

\begin{thebibliography}{1}
\expandafter\ifx\csname url\endcsname\relax
  \def\url#1{\texttt{#1}}\fi
\expandafter\ifx\csname urlprefix\endcsname\relax\def\urlprefix{URL }\fi

\bibitem{ewald}
P. Ewald, Ann. Phys {\bf 64}, 251 (1921)

\bibitem{FMM}
L. Greengard and V. Rokhlin, A fast algorithm for particle simulations,
J. Comput. Phys. 73, 325 (1987); 
H. Cheng, L. Greengard and V. Rokhlin, A Fast Adaptive Multipole
Algorithm in Three Dimensions, J. .Comput. Phys. 155, 468?498 (1999)

\bibitem{scuseria}
M.  Strain, G. Scuseria and M. Frisch, Science {\bf 271}, 51 (1996)
P. Maslen,  C. Ochsenfeld, C. White, M. Lee and M. Head-Gordon, J. Phys. Chem. {\bf 102}, 2215 (1998)
J. Perez-Jorda, and W. Yang, J. Chem. Phys. {\bf 107}, 1218 (1997)

\bibitem{Hockney.1970}
R.~W. Hockney, The potential calculations and some applications, Methods
  Comput. Phys. 9 (1970) 135--210.

\bibitem{daub} I. Daubechies, {\it ``Ten Lectures on Wavelets''}, SIAM, Philadelphia (1992) ; 
Goedecker, S., {\it Wavelets and their application for the solution of differential equations}, 1998b, Presses Polytechniques Universitaires et Romandes, Lausanne, Switzerland (ISBN 2-88074-398-2) 

\bibitem{oleg}
S. Goedecker, O. Ivanov, Linear Scaling solution of the classical Coulomb problem using
wavelets, Sol. State Comm., {\bf 105} 665 (1998)

\bibitem{lazy} G. Deslauriers and S. Dubuc, {\it Constr. Approx.} {\bf 5}, 49 (1989).

\bibitem{chelikowsky}
James R. Chelikowsky, N. Troullier, and Y. Saad, 
Finite-difference-pseudopotential method: Electronic structure calculations without a basis, 
Phys. Rev. Lett. 72, 1240 (1994)

\bibitem{pask}
J. E. Pask, B. M. Klein, C. Y. Fong, and P. A. Sterne, 
Real-space local polynomial basis for solid-state electronic-structure calculations: A finite-element approach, 
Phys. Rev. B 59, 12352 (1999) 

\bibitem{Martyna-Tuckerman.1999}
G.~J. Martyna, M.~E. Tuckerman, A reciprocal space based method for treating
  long range interactions in {\it ab initio} and force-field-based calculations
  in clusters, J. Chemical Physics 110~(6) (1999) 2810--2821.

\bibitem{Fusti-Molnar-Pulay.2002}
L.~F{\"u}sti-Molnar, P.~Pulay, Accurate molecular integrals and energies using
  combined plane wave and gaussian basis sets in molecular electronic structure
  theory, J. Chem. Phys. 116~(18) (2002) 7795--7805.

\bibitem{Beylkin-Mohlenkamp.2002}
G. Beylkin and L. Monzon,
Applied and Computational Harmonic Analysis, 19 (2005) 17-48 ; 

Algorithms for numerical analysis in high dimensions
 G. Beylkin and M. J. Mohlenkamp,
SIAM J. Sci. Comput.,  26 (6) (2005) 2133-2159; 

G.~Beylkin, M.~J. Mohlenkamp, Numerical operator calculus in higher dimensions,
  in: Proceedings of the National Academy of Sciences, Vol.~99, 2002, pp.
  10246--10251.

\bibitem{harrison}
The same expansion was used to find a wavelet representation of the 
entire integral operator in:
R. Harrison, G. Fann, T. Yanai, Z. Gan and G. Beylkin,
Multiresolution quantum chemistry: basic theory and initial applications
J. Chem. Phys. 121 (23) (2004) 11587-11598. 

\bibitem{hutter}
Ab initio molecular dynamics, 
D. Marx, J. Hutter, published in 'Modern methods and algorithms of quantum chemistry' 
J. Grotendorst editor, John von Neumann Institute for Computing, 2000


\bibitem{cpmd}
CPMD Version 3.3: developed by J. Hutter, A. Alavi, T. Deutsch,
M. Bernasconi, S. Goedecker, D. Marx, M. Tuckerman and M. Parrinello,
Max-Planck-Institut f\"{u}r Festk\"{o}rperforschung and IBM Z\"{u}rich Research Laboratory
(1995-1999)

\bibitem{beyl}  N. Saito, G. Beylkin, G.,
Signal Processing, IEEE Transactions on [see also Acoustics, Speech, and Signal Processing, IEEE Transactions on]
Vol. 41~(12),  (1993) 3584--3590



\end{thebibliography}
%\bibliographystyle{elsart-num}
%\begin{thebibliography}{00}

% \bibitem{label}
% Text of bibliographic item

% notes:
% \bibitem{label} \note

% subbibitems:
% \begin{subbibitems}{label}
% \bibitem{label1}
% \bibitem{label2}
% If there is a note, it should come last:
% \bibitem{label3} \note
% \end{subbibitems}

%\bibitem{}

%\end{thebibliography}

\end{document}